\documentclass[prl,superscriptaddress,twocolumn]{revtex4}
\usepackage{amsthm}
\usepackage{amsmath}
\usepackage{mathrsfs}
\usepackage{amssymb}
\usepackage{wasysym}
\usepackage{graphicx}
\usepackage{varwidth}
\usepackage{url}
\usepackage{dsfont}
\usepackage{float}
\usepackage{bm}
\usepackage{ifthen}
\usepackage[usenames,dvipsnames]{color}
\usepackage{mathrsfs}
\usepackage[colorlinks=true,citecolor=blue,urlcolor=black]{hyperref}
\usepackage{float}
\usepackage{afterpage}

\usepackage{algorithm, algorithmic}%
\usepackage{xcolor}
\newcommand{\sket}[1]{{\ensuremath{\lvert#1\rangle}}}
\newcommand{\lket}[1]{{\ensuremath{\left\lvert#1\right\rangle}}} 
\newcommand{\ket}[1]{\if@display\lket{#1}\else\sket{#1}\fi}
\newcommand{\sbra}[1]{{\ensuremath{\langle#1\rvert}}}
\newcommand{\lbra}[1]{{\ensuremath{\left\langle#1\right\rvert}}}
\newcommand{\bra}[1]{\if@display\lbra{#1}\else\sbra{#1}\fi}
\newcommand{\sbraket}[2]{{\ensuremath{\langle#1\rvert#2\rangle}}}
\newcommand{\lbraket}[2]{{\ensuremath{\left\langle#1\!\left\rvert\vphantom{#1}#2\right.\!\right\rangle}}}
\newcommand{\braket}[2]{\if@display\lbraket{#1}{#2}\else\sbraket{#1}{#2}\fi}

\newcommand{\sketbra}[2]{{\ensuremath{\lvert #1\rangle\!\langle #2\rvert}}}
\newcommand{\lketbra}[2]{{\ensuremath{\left\lvert #1\right\rangle\!\!\left\langle #2\right\rvert}}}
\newcommand{\ketbra}[2]{\if@display\lketbra{#1}{#2}\else\sketbra{#1}{#2}\fi}

\theoremstyle{plain}

\theoremstyle{definition}

\begin{document}

\title{Entangling Quantum Generative Adversarial Networks }

\author{Murphy Yuezhen Niu}
\thanks{These two authors contributed equally to this work.}
\affiliation{Google AI Quantum, Venice, CA 90291, USA}

\author{Alexander Zlokapa}
\thanks{These two authors contributed equally to this work.}
\affiliation{Division of Physics, Mathematics and Astronomy, Caltech, Pasadena, CA 91125, USA}
\affiliation{Google AI Quantum, Venice, CA 90291, USA}

\author{Michael Broughton}
\affiliation{Google AI Quantum, Venice, CA 90291, USA}

\author{Sergio Boixo}
\affiliation{Google AI Quantum, Venice, CA 90291, USA}

\author{Masoud Mohseni}
\affiliation{Google AI Quantum, Venice, CA 90291, USA}

\author{Vadim Smelyanskyi}
\affiliation{Google AI Quantum, Venice, CA 90291, USA}

\author{Hartmut Neven}
\affiliation{Google AI Quantum, Venice, CA 90291, USA}

\begin{abstract}
Generative adversarial networks~(GANs) are one of the most widely adopted semisupervised and unsupervised machine learning methods for high-definition image, video, and audio generation.
In this work, we propose a new type of architecture for quantum generative adversarial networks~(\emph{entangling} quantum GAN, EQ-GAN) that overcomes some limitations of previously proposed quantum GANs. Leveraging the entangling power of quantum circuits, EQ-GAN guarantees the convergence to a Nash equilibrium under minimax optimization of the discriminator and generator circuits by performing entangling operations between \textit{both} the generator output \textit{and} true quantum data. We show that EQ-GAN has additional robustness against coherent errors and demonstrate the effectiveness of EQ-GAN experimentally in a Google Sycamore superconducting quantum processor. By adversarially learning efficient representations of quantum states, we prepare an approximate quantum random access memory (QRAM) and demonstrate its use in applications including the  training of quantum neural networks.
\end{abstract}

\maketitle

\section{Introduction}%
Generative adversarial networks (GANs)~\cite{goodfellow2014generative} are one of the most widely adopted \emph{generative machine learning} methods, achieving state-of-the-art performance in a variety of high-dimensional and complex tasks including photorealistic image generation~\cite{NIPS2016_8a3363ab}, super-resolution~\cite{Ledig_2017_CVPR}, and molecular synthesis~\cite{molecule}. Given access only to a training dataset $S = \{x_i\}$ sampled from an underlying data distribution $p_\mathrm{data}(x)$, a GAN can generate realistic examples outside $S$. Certain probability distributions generated by   quantum computers are thought to be classically hard to sample from under plausible conjectures~\cite{boixo2016characterizing,bosonsampling,supremacy2019}, and learning to generate these samples using a classical GAN can also be formidably hard~\cite{niu2020learnability}. In this work, we focus on developing a fully quantum mechanical GAN, where the true data is given by a quantum state; the task is then to learn a \emph{generator} circuit that can reproduce the same quantum state. Following the framework of a GAN, a \emph{discriminator} circuit is presented either with the true data or with fake data from the generator. The generator and discriminator are then trained adversarially~\cite{szegedy2013intriguing}: the generator attempts to fool the discriminator, while the discriminator attempts to correctly distinguish true and fake data. While we focus on quantum data, we present viable applications of the resulting machine learning architecture in the context of classical data.

Recent work on a quantum GAN (QuGAN)~\cite{PhysRevA.98.012324,Lloyd2018} has proposed a direct analogy of the classical GAN architecture in designing the generator and discriminator circuits. We show that the proposed QuGAN does not always converge but rather in certain cases oscillates between a finite set of states due to mode collapse, and in general suffers from a non-unique Nash equilibrium.
This motivates a new \emph{entangling} quantum GAN (EQ-GAN) with a uniquely quantum twist: rather than providing the discriminator with \emph{either} true \emph{or} fake data, we allow the discriminator to entangle \emph{both} true and fake data. We prove the convergence of the EQ-GAN to the global optimal Nash equilibrium. Numerical experiments confirm that the EQ-GAN converges on problem instances that the QuGAN failed on.

While the EQ-GAN has favorable convergence properties, the task of learning a quantum circuit to generate an unknown quantum state may also be solved in an entirely supervised approach. Rather than adversarially training the discriminator to distinguish between fake and real data, one may freeze the discriminator to perform an exact swap test, measuring the state fidelity between the true and fake data. While this would replicate the original state in the absence of noise, gate errors in the implementation of the discriminator will cause convergence to the incorrect optimum. We show that the adversarial approach of the EQ-GAN is more robust to such errors than the simpler supervised learning approach.  Since training quantum machine learning models can require extensive time to compute gradients on current quantum hardware, resilience to gate errors drifting during the training process is especially valuable in the noisy intermediate-scale quantum (NISQ) era of quantum computing.

Finally, we provide applications of the EQ-GAN in the broader context of quantum machine learning for \emph{classical} data. Many of the most attractive quantum machine learning algorithms require a quantum random access memory (QRAM)~\cite{lloyd2016}. By learning a shallow quantum circuit to generate a superposition of classical data, an EQ-GAN can be used to create an approximate QRAM. We demonstrate an application of such a QRAM for quantum neural networks~\cite{farhi2018classification}, improving the performance of a quantum neural network for a classification task.

\section{Prior Art}%
A GAN comprises of a parameterized generative network $G(\theta_g, z)$ and discriminator network $D(\theta_d, z)$. The generator maps a vector sampled from an input distribution $z \sim p_0(z)$ to a data example $G(\theta_g, z)$, thus transforming $p_0(z)$ to a new distribution $p_g(z)$ of fake data. The discriminator takes an input sample $x$ and gives the probability $D(\theta_d, x)$ that the sample is real (from the data) or fake (from the generative network). The training corresponds to a minimax optimization problem, where we alternate between improving the discriminator's ability to distinguish real/fake samples and improving the generator's ability to fool the discriminator. Specifically, we solve $\min_{\mathbf{\theta}_g}\max_{\mathbf{\theta}_d} V(\theta_g, \theta_d)$ for a cost function $V$:

\begin{align}
\begin{split}
\label{eq:classical}
  V(\theta_g, \theta_d) &= \mathbb{E}_{x\sim p_\mathrm{data}(x)} \left[\log D(\mathbf{\theta}_d, x) \right] \\
  				  &\quad+ \mathbb{E}_{z\sim p_0(z)} \left[\log\left(1-D(\mathbf{\theta}_d, G(\theta_g, z))\right)\right].
\end{split}
\end{align}

If $G$ and $D$ have enough capacity, i.e. approach the space of arbitrary functions, then it is proven in Ref.~\cite{goodfellow2014generative} that the global optimum of this minimax game exists and uniquely corresponds to $p_g(x) = p_\mathrm{data}(x)$. While a multilayer perceptron can be used to parameterize $D$ and $G$, the dimensionality of the functional space can also be increased by replacing classical neural networks with quantum neural networks. In the most general case, the classical data can be represented by a density matrix $\sigma=\sum_i p_i \ket{\psi_i}\bra{\psi_i}$ where $p_i \in [0,1]$ are positive bounded real numbers and $\ket{\psi_i}$ are orthogonal basis states. In the first proposal of a quantum GAN (QuGAN)~\cite{PhysRevA.98.012324,Lloyd2018}, the generative network is defined by a quantum circuit $U$ that outputs the quantum state $\rho=U(\theta_g)\rho_0 U^\dagger(\theta_g)$ from the initial state $\rho_0$. The discriminator takes either the real data $\sigma$ or the fake data $\rho$ and performs a positive operator valued measurement (POVM) defined by    $T$ whose outcome determines the probability of data being true,  or operator $F$ whose outcome determines the probability of data being fake, with $||T||_1, ||F||_1 \leq 1$. Hence, the discriminator predicts the probability that an unknown input state $\rho_\mathrm{in}$ is true data by measuring the expectation value of $T$:
\begin{align}
    D(\theta_d, \rho_\mathrm{in}) = \text{Tr}[T \rho_\mathrm{in}].
\end{align}
Following Ref.~\cite{Lloyd2018}, the QuGAN solves the minimax game
\begin{align}
\label{minmaxQuGANsEq}
    \min_{\mathbf{\theta}_g}\max_{T}\left( \text{Tr}[T
    \sigma] - \text{Tr}[T
    \rho(\mathbf{\theta}_g)]\right)\;.
\end{align} 

Unfortunately, minimax optimization might not converge to a good Nash equilibrium. When $\rho$  is close to $\sigma$, the optimal Hesltrom measurement operator $T = P^+(\sigma-\rho)$ is close to orthogonal to the true quantum data $\sigma$ and opposite to $\rho$. The next step of training will try align the generator state $\rho$ with $T$ to minimize the cost function in Eq.~\ref{minmaxQuGANsEq}, perhaps overshooting $\sigma$. In the subsequent generator update, $T$ will again be opposite to $\rho$. This leads to the oscillation of the generator and discriminator, possibly preventing convergence; we show a case of infinite oscillation in the supplementary material.
 

\section{Convergence of EQ-GAN}%
\begin{figure}[h!]
\begin{center}
\includegraphics[width=0.9\linewidth]{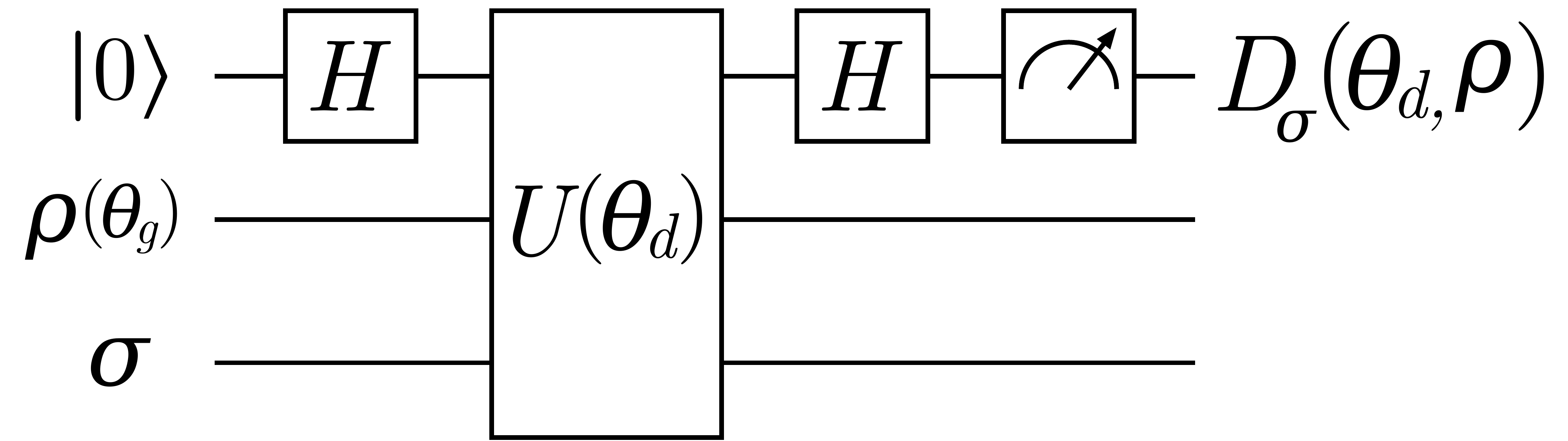}

\caption{EQ-GAN architecture. The generator varies $\theta_g$ to fool the discriminator; the discriminator varies $\theta_d$ to distinguish the state. Since an optimal discriminator performs a swap test, the global optimum of the EQ-GAN occurs when $\rho(\theta_g) = \sigma$. While we include an ancilla qubit in the figure for clarity, we implement a destructive ancilla-free swap test~\cite{PhysRevA.87.052330}.}
\label{fig:eqgan}
\end{center}
\end{figure}

To ensure convergence to a unique Nash equilibrium, we propose a new minimax optimization problem with a discriminator that is not directly analogous to the discriminator of a classical GAN. Rather than evaluating either fake or true data individually, the optimal discriminator is not only provided access to the true data $\sigma$ and an input state $\rho(\theta_g)$ prepared by the generator circuit parameterized by $\theta_g$,  but also permited to perform a measurement on the joint system that in certain parameter value gives fidelity measurement between the two inputs:
\begin{align}
D_\sigma^\mathrm{fid}(\rho(\theta_g)) = \left(\text{Tr}\sqrt{ \sigma^{1/2} \, \rho(\theta_g) \, \sigma^{1/2}} \, \right)^2.
\end{align} Notice that in comparison Eq.~\ref{minmaxQuGANsEq} is a linear function of input states, which is not optimal in the state-certification problem~\cite{buadescu2019quantum} of evaluating quantum generative models.
Let the discriminator $D_\sigma(\theta_d, \rho(\theta_g))$ represent the probability of measuring state $\ket{0}$ at the end of the discriminating circuit. If there exist parameters $\theta_d^\mathrm{opt}$ that realize a perfect swap test, i.e. $D_\sigma(\theta_d^\mathrm{opt}, \rho(\theta_g)) = \frac{1}{2} + \frac{1}{2}D_\sigma^\mathrm{fid}(\rho(\theta_g))$, then $D_\sigma$ is sufficiently expressive to reach the optimal discriminator during optimization. 
Since a traditional swap test across two $n$-qubit states requires two-qubit gates that span over $2n$ qubits, implementation on a quantum device with local connectivity incurs prohibitive overhead in circuit depth. Hence, we implement the discriminator with a parameterized destructive ancilla-free swap test~\cite{PhysRevA.87.052330}. The EQ-GAN architecture adversarially optimizes the generation of the state $\rho(\theta_g)$ and the learning of a fidelity measurement $D_\sigma$ (Fig.~\ref{fig:eqgan}).

We define a minimax cost function closer to that of the classical GAN in Eq.~\ref{eq:classical}:

\begin{align}
\begin{split}
\label{eq:eqgan}
  \min_{\mathbf{\theta}_g}\max_{\mathbf{\theta}_d} V(\theta_g, \theta_d)  &=\min_{\mathbf{\theta}_g}\max_{\mathbf{\theta}_d} [1-  D_\sigma(\mathbf{\theta}_d, \rho(\theta_g))],
\end{split}
\end{align}
where $D_\sigma(\mathbf{\theta}_d, \rho(\theta_g))$ is the parameterization of the swap-test result.
We now show that a \emph{Nash equilibrium exists} at the desired location.  Consider a swap test circuit ansatz for the discriminator $U(\theta_d)=\exp[-i\theta_d\text{CSWAP}]$, which is the matrix exponentiation of a perfect controlled swap gate with angle $\theta_d$. Under such ansatz, the  input state $\rho_{\text{in}}=\ket{\psi}\bra{\psi}$ and $\sigma=\ket{\zeta}\bra{\zeta}$ will transform under the discriminator circuit into:
\begin{widetext}
\begin{align}\label{eq:discriminatorcircuit}
&H U(\theta_d)H\ket{0}_\text{a}\ket{\psi}\ket{\zeta}=\frac{i\sin\theta_d}{2} \ket{1}_{\text{a}}[\ket{\zeta}\ket{\psi}-\ket{\psi}\ket{\zeta}]  + \frac{1}{2}\ket{0}_{\text{a}}[(e^{-i\theta_d} +\cos\theta_d)  \ket{\psi}\ket{\zeta} - i \sin\theta_d\ket{\zeta}\ket{\psi}].
\end{align} 
\end{widetext}
Given the circuit ansatz defined above  with the predefined range for the swap angle $\theta$, the maximum value for distinguishing between two arbitrary states is uniquely achieved by perfect swap test angle $\theta=\pi/2$. More particularly, the probability of measuring state $\ket{0}$ at the end of the parameterized swap test depends on the swap angle $\theta$ according to
\begin{align}\label{eqpzero}
   D_\sigma(\mathbf{\theta}_d, \rho(\theta_g)) =\frac{1}{2}[1+\cos^2\theta_d + \sin^2\theta_d D_\sigma^\mathrm{fid}(\rho(\theta_g))].
\end{align}
The discriminator aims to decrease the probablity of measuring $\ket{0}$, and thus minimize Eq.~\ref{eqpzero} by getting close to $\theta_d =\pi/2$ which corresponds to the perfect swap test given $D_\sigma^\mathrm{fid}(\rho(\theta_g)))\leq 1$. The next step is for the generator to maximize $D_\sigma^\mathrm{fid}(\rho(\theta_g)))$ by moving closer to the true data.
Ultimately, the generator cannot improve when $\rho(\theta_g) = \sigma$.

The cost function defined in Eq.~(\ref{eq:eqgan}) does not assume that the input states $\sigma$ and $\rho(\theta_g)$ have to be pure states. For simplicity, the example we provided in Eq.~(\ref{eq:discriminatorcircuit}) does assume pure state input. In the supplementary material, we discuss one proposal for EQ-GAN to accommodate mixed state input by replacing the pure-state fidelity with a mixed state fidelity measurement. Moreover, other discriminator architectures may be chosen to ensure the existence of a Nash equilibrium. In the experiments presented below, we use a hardware-efficient ansatz designed to correct dominant coherent gate errors. Although a poorly chosen circuit parameterization may yield a non-convex loss function landscape and thus be difficult to optimize by gradient descent, this is an issue shared with the QuGAN due to the difficulty of expressing arbitrary unitaries as shallow quantum circuits as well as with classical GANs. However, the EQ-GAN architecture successfully converges on problem instances that are unreachable by a fully trained and properly parameterized a QuGAN (see supplementary material).

\section{Learning to suppress errors}%
We now show the improved robustness of an EQ-GAN against gate errors compared to a more straightforward swap test approach to learning an unknown quantum state. Rather than adversarially training the parameterized swap test used as a discriminator in EQ-GAN, a perfect swap test could be applied every iteration by a frozen discriminator. This may also cause the generator circuit to converge to the true data, since the swap test ensures a unique global optimum.

However, in the presence of gate errors in the swap test, this unique global optimum will be offset from the true data. Since EQ-GAN is agnostic to the precise parameterization of a perfect swap test, an appropriate ansatz can learn to correct coherent errors observed on near-term quantum hardware. In particular, the gate parameters such as conditional $Z$ phase, single qubit $Z$ phase and swap angles in two-qubit entangling gate can drift and oscillate over the time scale of $O(10)$ minutes~\cite{arute2020supp,arute2020observation}. Such unknown systematic and time-dependent coherent errors provides significant challenges for applications in quantum machine learning where gradient computation and update requires many measurements.


The large deviations in single-qubit and two-qubit $Z$ rotation angles can largely be mitigated by including additional single-qubit $Z$ phase compensations. The effectiveness and importance of such systematic error mitigation is recently demonstrated in the success of achieving the-state-of-art accuracy in energy estimation for fermionic molecules~\cite{HartreeFockScience}. In learning the discriminator circuit that is closest to a true swap test, the adversarial learning of EQ-GAN provides a useful paradigm that may be broadly applicable to improving the fidelity of other near-term quantum algorithms.

Suppose the adversarial discriminator unitary is given by $U(\theta_d)$, where $U(\theta_d^\mathrm{opt})$ corresponds to a perfect swap test in the absence of noise. Given a trace-preserving completely positive noisy channel $\mathcal{E}$, the discriminator is replaced by a new unitary operation $\tilde U(\theta_d)$. While a supervised approach would apply an approximate swap test given by $\tilde U(\theta_d^\mathrm{opt})$, the adversarial swap test will generically perform better if there exist parameters $\theta^*_d$ such that $||\tilde U(\theta^*_d) -  U(\theta_d^\mathrm{opt})||_2 < || \tilde U(\theta_d^\mathrm{opt}) - U(\theta_d^\mathrm{opt})||_2$. Because the discriminator defines the loss landscape optimized by the generator, the $\rho(\theta_g)$ produced by EQ-GAN may converge to a state closer to $\sigma$ than possible by a supervised approach if the parameterization of the noisy unitary $\tilde U$ is general enough to mitigate errors.

\begin{figure}[h!]
\begin{center}
\includegraphics[width=0.79\linewidth]{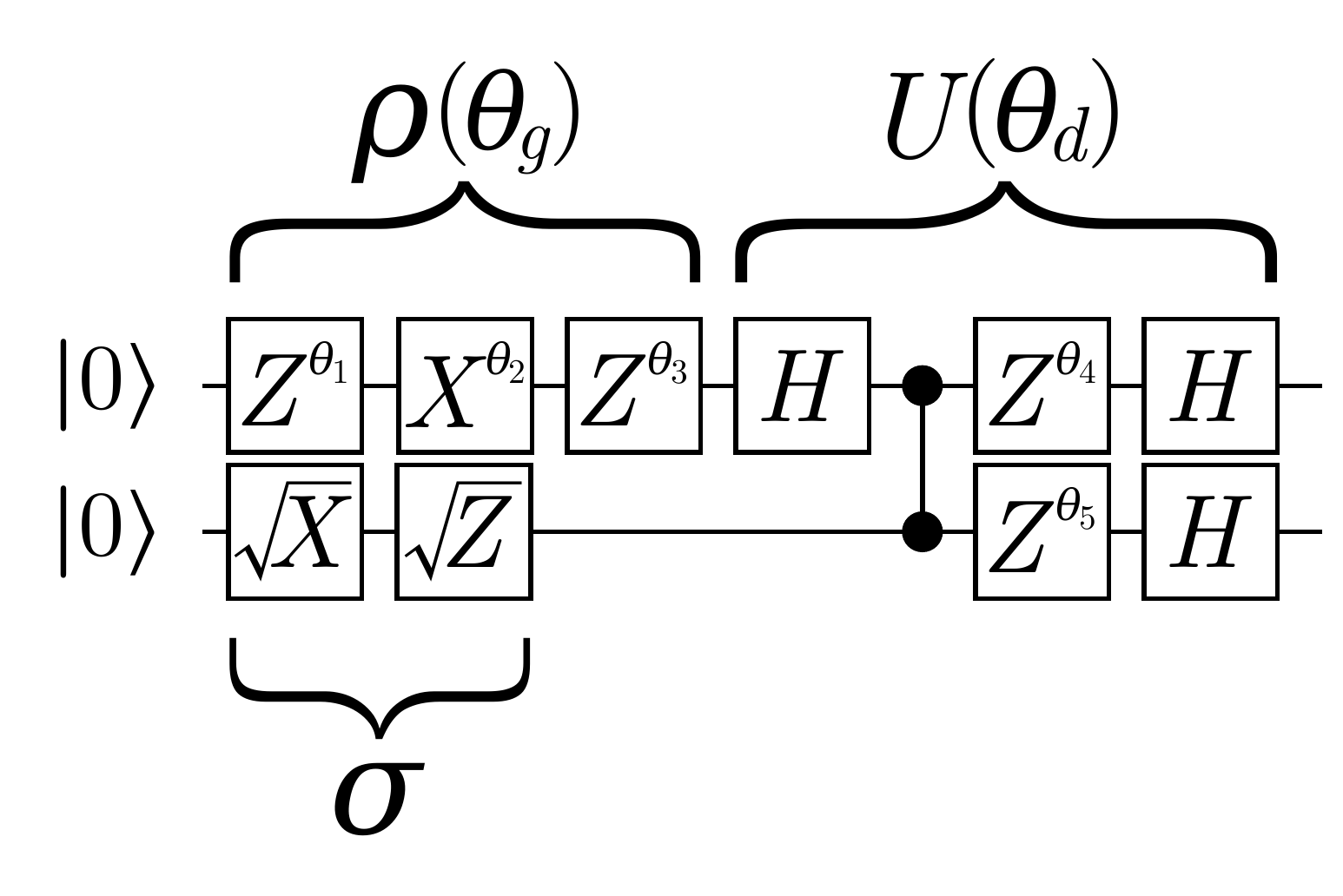}
\caption{EQ-GAN experiment for learning a single-qubit state. The discriminator $(U(\theta_d)$ is constructed with free $Z$ rotation angles to suppress CZ gate errors, allowing the generator $\rho(\theta_g)$ to converge closer to the true data state $\sigma$ by varying $X$ and $Z$ rotation angles.}
\label{fig:experiment}
\end{center}
\end{figure}

\begin{figure}[h!]
\begin{center}
\includegraphics[width=0.9\linewidth]{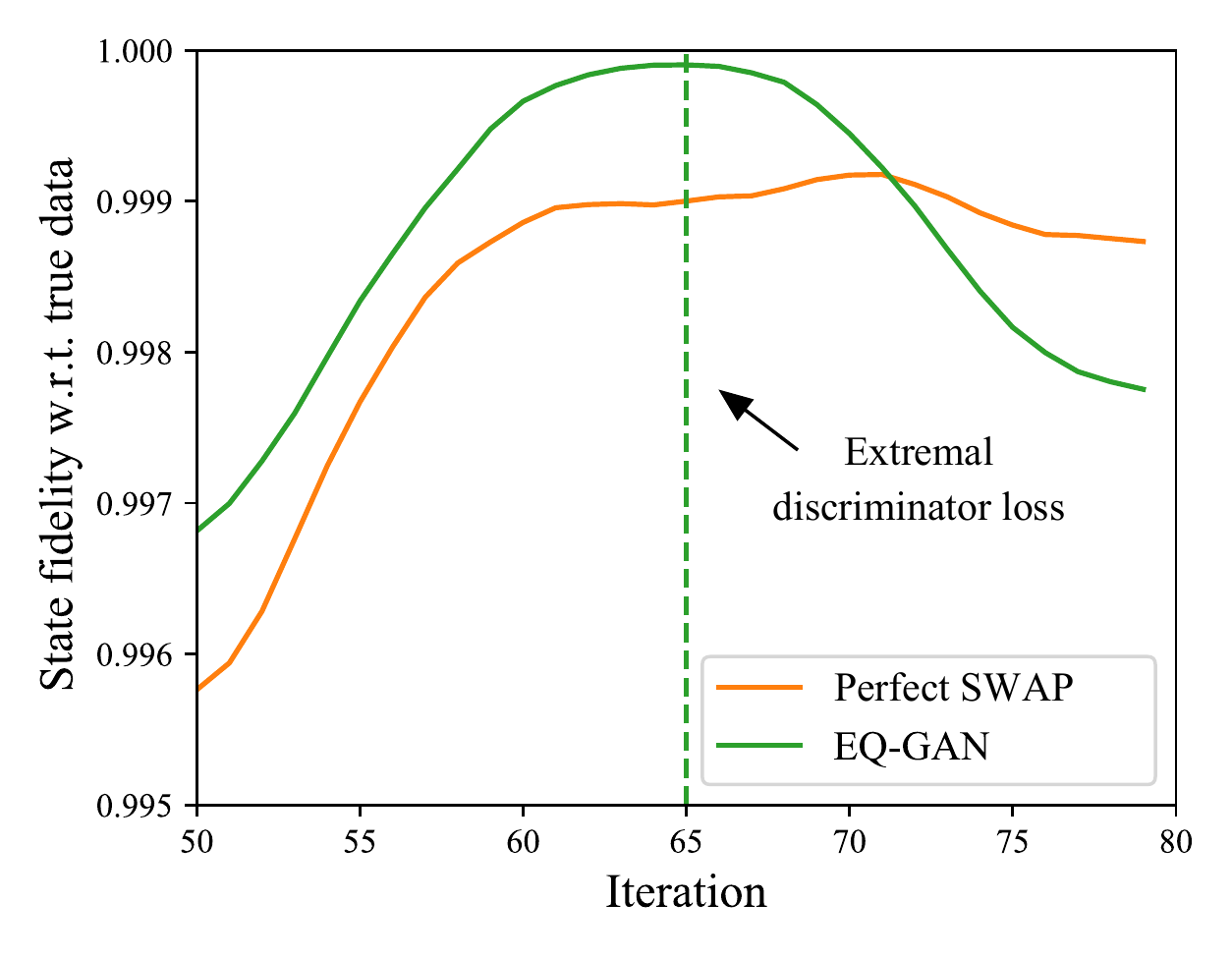}
\caption{Comparison of EQ-GAN and a perfect swap test on a physical quantum device. We experimentally confirm that the EQ-GAN converges to a higher state overlap by learning to correct such errors with additional single-qubit rotations. The ``converged'' EQ-GAN (dashed line) coincides with the iteration where the discriminator loss is minimized.}
\label{fig:err}
\end{center}
\end{figure}

As an example, we consider the task of learning the superposition $\frac{1}{\sqrt{2}}(\ket{0} + \ket{1})$ on a quantum device with noise (Fig.~\ref{fig:experiment}). The discriminator is defined by a swap test with CZ gate providing the necessary two-qubit operation. To learn to correct gate errors, however, the discriminator adversarially learns the angles of single-qubit $Z$ rotations insert directly after the CZ gate. Hence, the EQ-GAN obtains a state overlap significantly better than that of the perfect swap test (Fig.~\ref{fig:err}). Although both methods do not stay at the optimal point, this is typical of noisy gradient measurements and minimax optimization:after convergence to the Nash equilibrium, discretization can induce perturbations while non-zero higher-order gradients lead the training to deviate from the global optimum~\cite{brock2018large}.

We report the average error after multiple runs of the EQ-GAN and perfect swap test on an experimental device (Table~\ref{tab:err}).

\begin{table}[h!]
\centering
\begin{tabular}{ p{3cm}||p{4.5cm} }
 QML model & Minimum error in state fidelity\\
 \hline
 Perfect swap & $(2.4 \pm 0.5) \times 10^{-4}$ \\
 \textbf{EQ-GAN} & $(\mathbf{0.6 \pm 0.2) \times 10^{-4}}$
  \end{tabular} 
 \caption{Comparison of EQ-GAN and a perfect swap test on a Sycamore quantum device. The error of the EQ-GAN (i.e. $1 -\, \mathrm{state\; fidelity}$) is significantly lower than that of the perfect swap test, demonstrating the successful adversarial training of an error-suppressed swap test. Uncertainties show two standard deviations.}
 \label{tab:err}
\end{table}

\section{Application to QRAM}%
Many quantum machine learning applications require a quantum random access memory (QRAM) to load \emph{classical} data in superposition~\cite{lloyd2016}. More particularly, a set of classical data can be described by the empirical distribution $\{P_i\}$ over all possible input data $i$. Most quantum machine learning algorithms require the conversion from  $\{P_i\}$ into a quantum state $\sum_i \sqrt{P_i}\ket{\psi_i}$, i.e. a superposition of orthogonal basis states $\ket{\psi_i}$ representing each single classical data entry with an amplitude proportional to the square root of the classical probability $P_i$. Preparing such a superposition of an arbitrary set of $n$ states takes $O(n)$ operations at best, which ruins the exponential speedup. Given a suitable ansatz, we may use an EQ-GAN to learn a state approximately equivalent to the superposition of data.  

\begin{figure}[h!]
\begin{center}
\includegraphics[width=\linewidth]{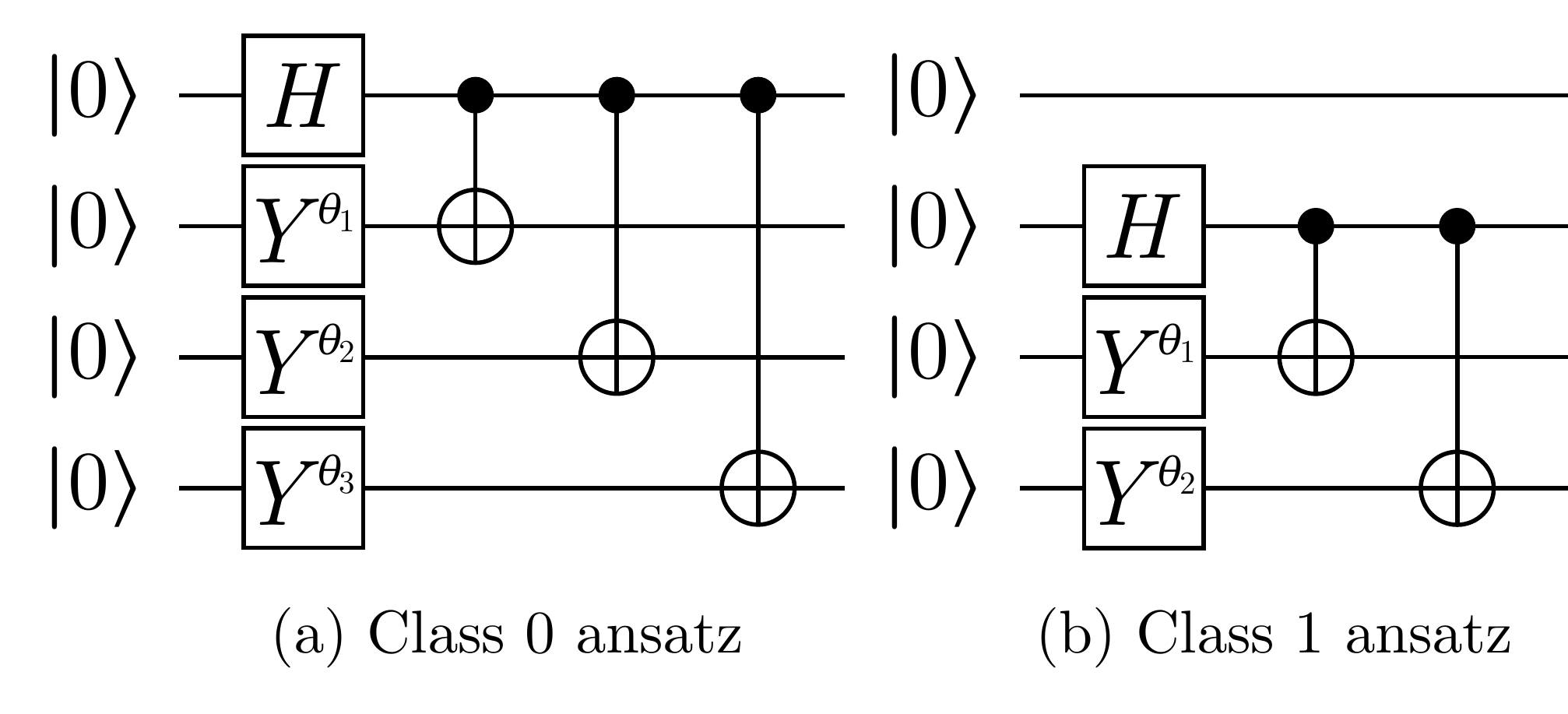}
\caption{Variational QRAM ansatzes for generating peaks by learning $\theta_i$ parameters~\cite{PhysRevA.102.012612}. Class 0 corresponds to a centered peak, and Class 1 corresponds to an offset peak.}
\label{fig:pansatz}
\end{center}
\end{figure}

\begin{figure}[h!]
\begin{center}
\includegraphics[width=\linewidth]{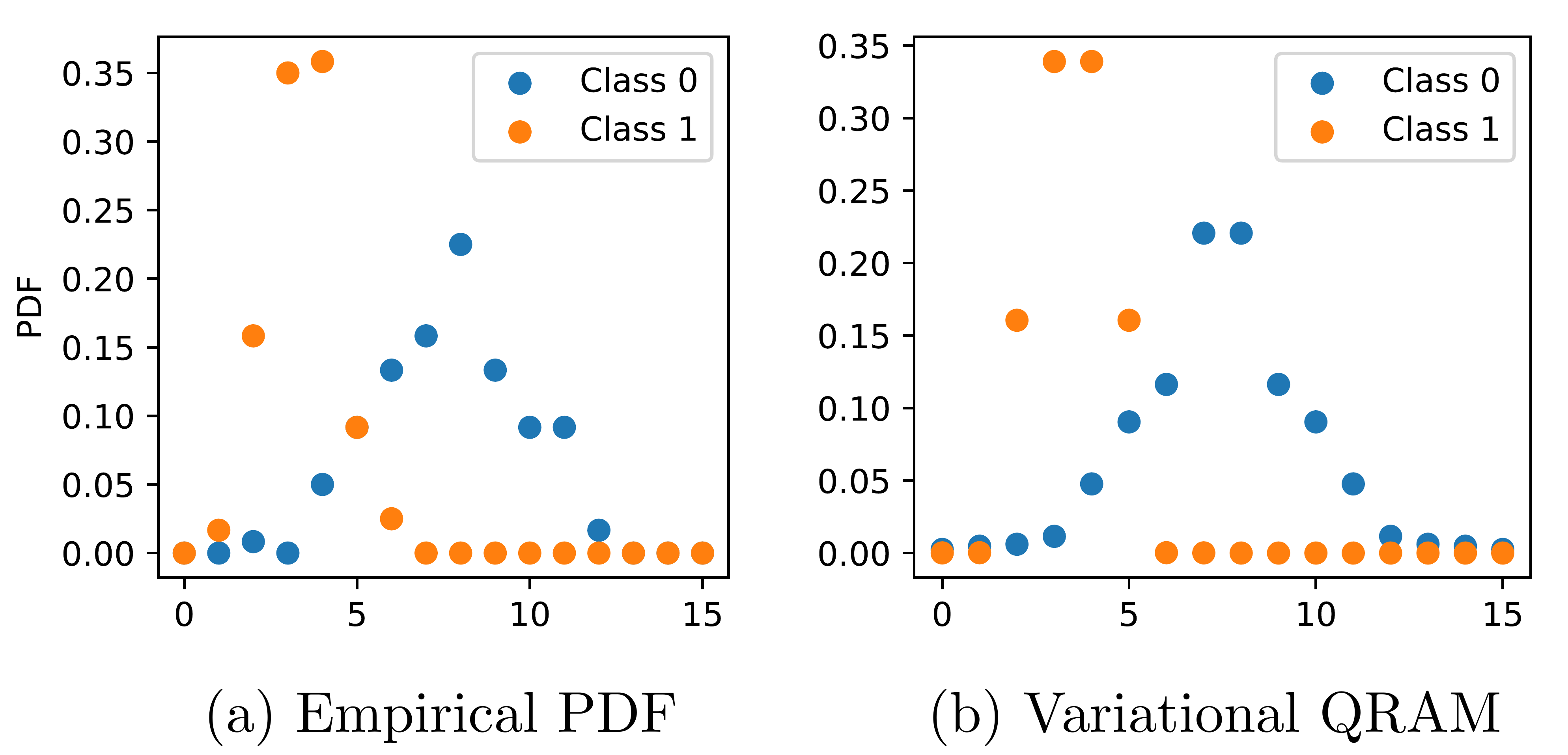}
\caption{Two-peak total dataset (sampled from normal distributions, $N=120$) and variational QRAM of the training dataset ($N=60$). The variational QRAM is obtained by training an EQ-GAN to generate a state $\rho$ with the shallow peak ansatz to approximate an exact superposition of states $\sigma$. The training and test datasets (each $N=60$) are both balanced between the two classes.}
\label{fig:pencode}
\end{center}
\end{figure} 

To demonstrate a variational QRAM, we consider a dataset of two peaks sampled from different Gaussian distributions. Exactly encoding the empirical probability density function requires a very deep circuit and multiple-control rotations; similarly, preparing a Gaussian distribution on a device with planar connectivity requires deep circuits (see supplementary material). Hence, we select shallow circuit ansatzes (Fig.~\ref{fig:pansatz}) that generate concatenated exponential functions to approximate a symmetric peak~\cite{PhysRevA.102.012612}. Once trained to approximate the empirical data distribution, the variational QRAM closely reproduces the original dataset (Fig.~\ref{fig:pencode}).

\begin{table}[h!]
\centering
\begin{tabular}{ p{4cm}||p{2cm} }
 Training data & Accuracy\\
 \hline
 Exact sampling & $53\% \pm 6\%$ \\
 \textbf{Variational QRAM} & $\mathbf{69\% \pm 2\%}$
  \end{tabular} 
 \caption{Test accuracy ($N=60$) of a quantum neural network (QNN) either trained on the all samples of the training dataset ($N=60$) for a single epoch or trained on the variational QRAM for an equal number of circuit evaluations. Although the QNN trained on the variational QRAM did not have direct access to the original dataset, accuracy is evaluated on the raw dataset. Uncertainties show two standard deviations.}
 \label{tab:res}
\end{table}

As a proof of principle for using such QRAM in a quantum machine learning context, we train a quantum neural network~\cite{farhi2018classification} and compute hinge loss either by considering each data entry individually (encoded as a quantum circuit) or by considering each class individually (encoded as a superposition in variational QRAM). Given the same number of circuit evaluations to compute gradients, the superposition converges to a better accuracy at the end of training despite using an approximate distribution (Table~\ref{tab:res}).

\section{Conclusion}
Motivated by limitations of preexisting quantum GAN architectures in the literature, we propose the EQ-GAN architecture to overcome issues of non-convexity and mode collapse. We adopt a parameterization of Hilbert-Schmidt norm as the cost function as oppose to trace distance based on the optimality of Hilbert-Scmidt norm in state-certification problems. Similar advantages of Hilbert-Schmidt norm has been shown in quantum embedding designs of quantum kernel learning~\cite{lloyd2020quantum}.  Other approaches to a quantum GAN may improve a quantum GAN's convergence properties --- notably, recent work suggests that certain cost functions such as the Wasserstein metric may provide more robust convergence~\cite{kiani2021quantum}. However, we find that the EQ-GAN's shallow discriminator is effective at suppressing device errors and ensures robust convergence in running laboratory quantum computers, making the EQ-GAN particularly relevant for near-term applications of quantum computing.  Moreover, we demonstrate the first experimental applicaiton of EQ-GAN using Google's cloud quantum computers in improving the classification accuracy of classical data using QRAM.  This work opens up new directions in utilizing quantum generative models to achieve  quantum speedup in  machine learning that necessitates the efficient and high-delity preparation of QRAM.

Open source code of this paper is available in GitHub
repository~\footnote{\url{https://github.com/tensorflow/quantum/tree/research/eq_gan}}.

\section{Acknowledgement}
AZ acknowledges support from Caltech's Intelligent Quantum Networks and Technologies (INQNET) research program and by the DOE/HEP QuantISED program grant, Quantum Machine Learning and Quantum Computation Frameworks (QMLQCF) for HEP, award number DE-SC0019227.

\clearpage
\appendix
\section{Mode Collapse Example of QuGAN}
We provide a concrete example of mode collapse in the original QuGAN architecture~\cite{PhysRevA.98.012324,Lloyd2018}. Consider a true data state $\sigma$ and a generator initialized in state $\rho$, where each state is defined by
\begin{align}
\label{modecollapseTargetEq}
\sigma = \frac{1 + \cos(\pi/6)\sigma^x + \sin(\pi/6)\sigma^y}{2},\\
\label{modecollapseGeneratorEq}
    \rho=\frac{1 + \cos(\pi/6)\sigma^x - \sin(\pi/6)\sigma^y}{2}.
\end{align}

\begin{figure}[h!]
\begin{center}
\includegraphics[width=\linewidth]{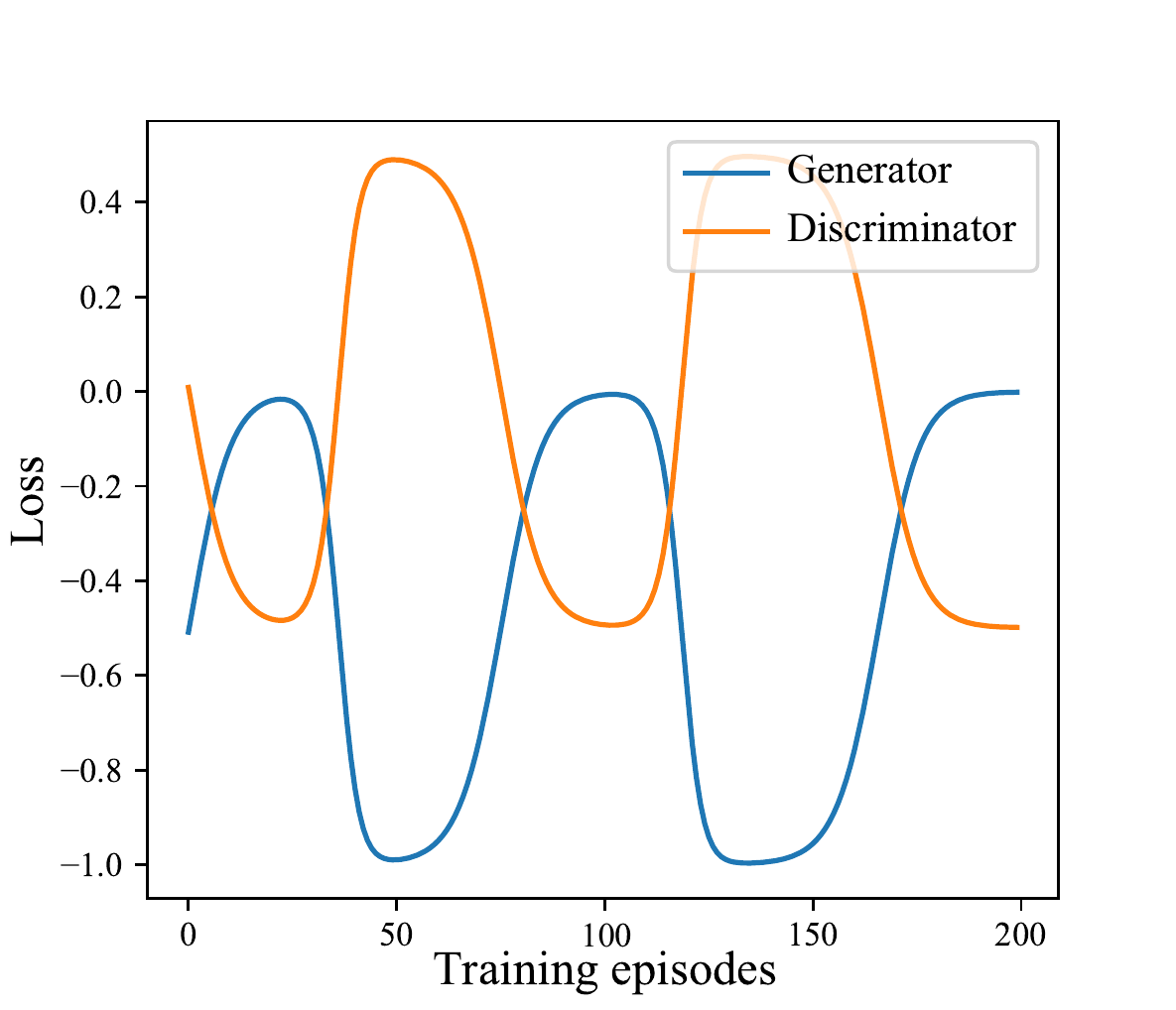}
\caption{Performance of QuGAN~\cite{PhysRevA.98.012324,Lloyd2018} learning the state defined in Eq.~(\ref{modecollapseTargetEq}) with initialization given by Eq.~\ref{modecollapseGeneratorEq}. Mode collapse manifests as an oscillation in the generator and discriminator loss without converging to a global optimum. The implementation is based on the original architecture in Pennylane~\cite{bergholm2020pennylane}.}
\label{fig:pennylane}
\end{center}
\end{figure}

Maximizing Eq.~\ref{minmaxQuGANsEq} with a Helstrom measurement by decomposing $\sigma - \rho = \frac{\sigma^y}{2}$, the discriminator will take $T = P^+(\sigma-\rho) = \frac{1+\sigma^y}{2}$. Optimizing over the space of density matrices, the generator will rotate $\rho$ to be parallel to $T$, also giving $\rho' = \frac{1+\sigma^y}{2}$. In the next iteration, the discriminator attempts to perform a new Helstrom measurement to distinguish $\sigma$ from $\rho'$, but this results in $T' = P^+(\sigma - \rho') = \rho$. As the generator realigns to match the new measurement operator, we find that $\rho'' = \rho$. It is now straightforward to see that if the QuGAN is trained to fully solve the minimax optimization problem each iteration, it will never converge; instead, it will always oscillate between states $\rho$ and $\rho'$, neither of which are the Nash equilibrium of the minimax game (Fig.~\ref{fig:pennylane}) for the QuGAN performance under such mode-collapse.

More generally, we can consider the oscillation between a finite set of states. Let the function $T_\sigma(\rho) = P^+(\sigma - \rho)$ denote the optimal Helstrom measurement $P^+ = \sum_i\ket{\phi_i^+}\bra{\phi_i^+}$ obtained from the positive part of the spectral decomposition of $\sigma - \rho$. If $T_\sigma^{(k)}$ is the $k$-fold composition of $T$ with itself, then the existence of some $k > 1$ such that $T^{(k)} = \rho$ is sufficient to ensure oscillation between $k$ states. For a system of $n$ qubits, we may achieve this by preparing the target and initial state separated by an angle of $\pi/3$ on the generalized Bloch sphere.

\section{Training EQ-GAN}
While the original QuGAN architecture is shown to oscillate indefinitely for an example constructed in the main text (see Fig.~\ref{fig:pennylane}), we provide numerical experiments here to demonstrate the successful convergence of the proposed EQ-GAN architecture.

\begin{figure}[h!]
\begin{center}
\includegraphics[width=\linewidth]{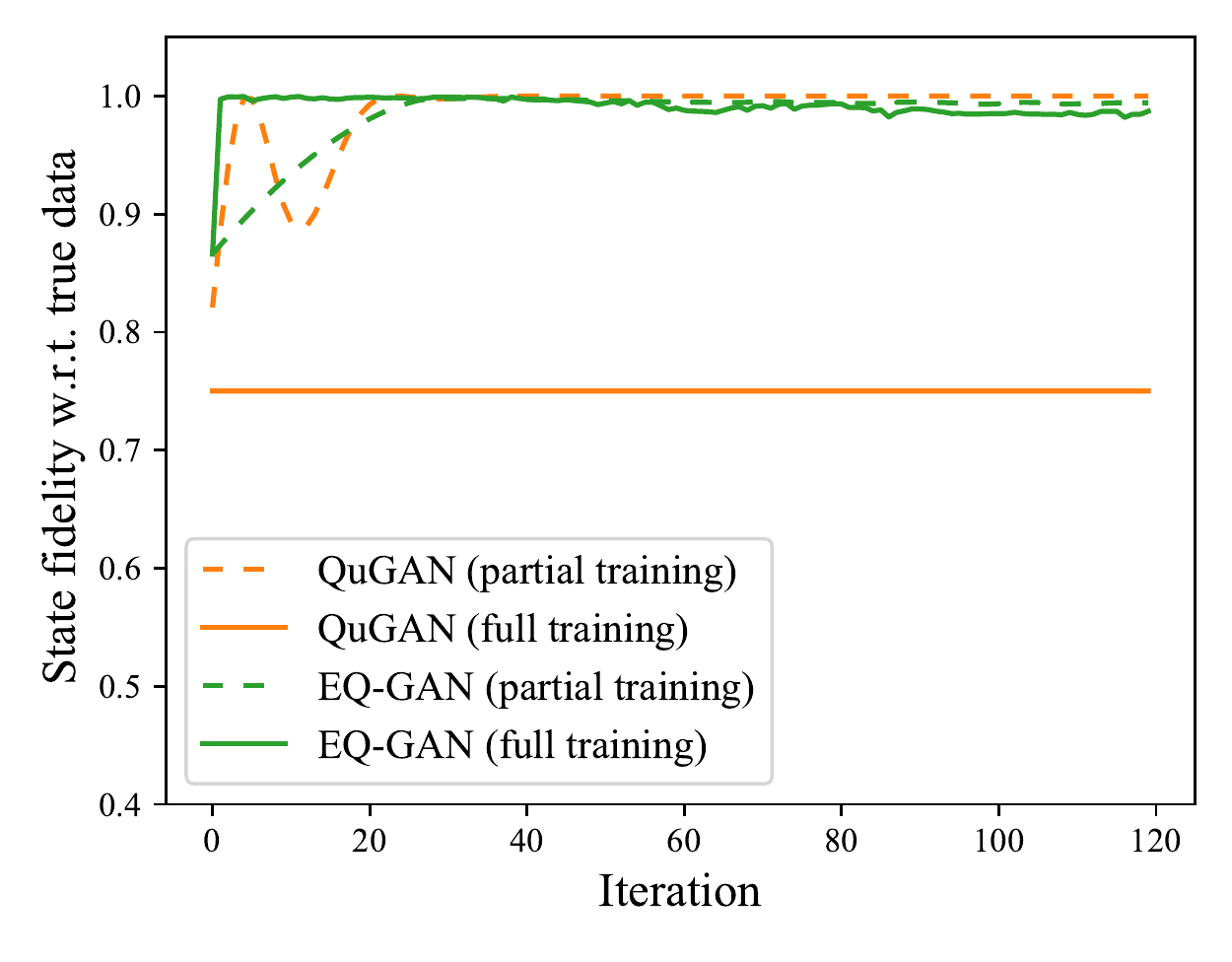}
\caption{Comparison of QuGAN~\cite{PhysRevA.98.012324,Lloyd2018} and EQ-GAN learning the state given by Eq.~\ref{modecollapseTargetEq}. \emph{Full training} denotes training the generator for 50 epochs then the discriminator for 50 epochs each iteration; \emph{partial training} denotes only 1 epoch per iteration. The QuGAN remains more unstable than EQ-GAN during training with either training configuration.}
\label{fig:osc}
\end{center}
\end{figure}

In Fig.~\ref{fig:osc}, we illustrate a subtlety in the oscillatory analysis presented in the main text. Within the GAN formalism, the generator and discriminator iteratively optimize a given loss function. When the optimization is allowed to converge to an extremum of the loss function in the QuGAN architecture specifically, the result is determined by a Helstrom measurement. It is for this case that indefinite oscillation is shown; in the case of learning the state $\sigma$ constructed in the main text, oscillation between states $\rho$ and $\rho'$ result in a constant state overlap of $3/4$.

However, the iterative optimization procedure to move towards the optimal Helstrom measurement may be only partly completed, i.e. the generator and discriminator are not allowed to extremize the loss function. With such a selection of hyperparameters, we observe that oscillation between states continues (Fig.~\ref{fig:osc}), leading to unstable training for the QuGAN architecture. In comparison, the same hyperparameters perform well for the EQ-GAN architecture, which steadily approaches the true data state. Unstable training is difficult to overcome even in classical GAN architectures~\cite{pmlr-v80-mescheder18a}, and thus advances in understanding how to prevent such non-convergence are consequential for both quantum and classical machine learning.

\begin{figure}[h!]
\begin{center}
\includegraphics[width=0.95\linewidth]{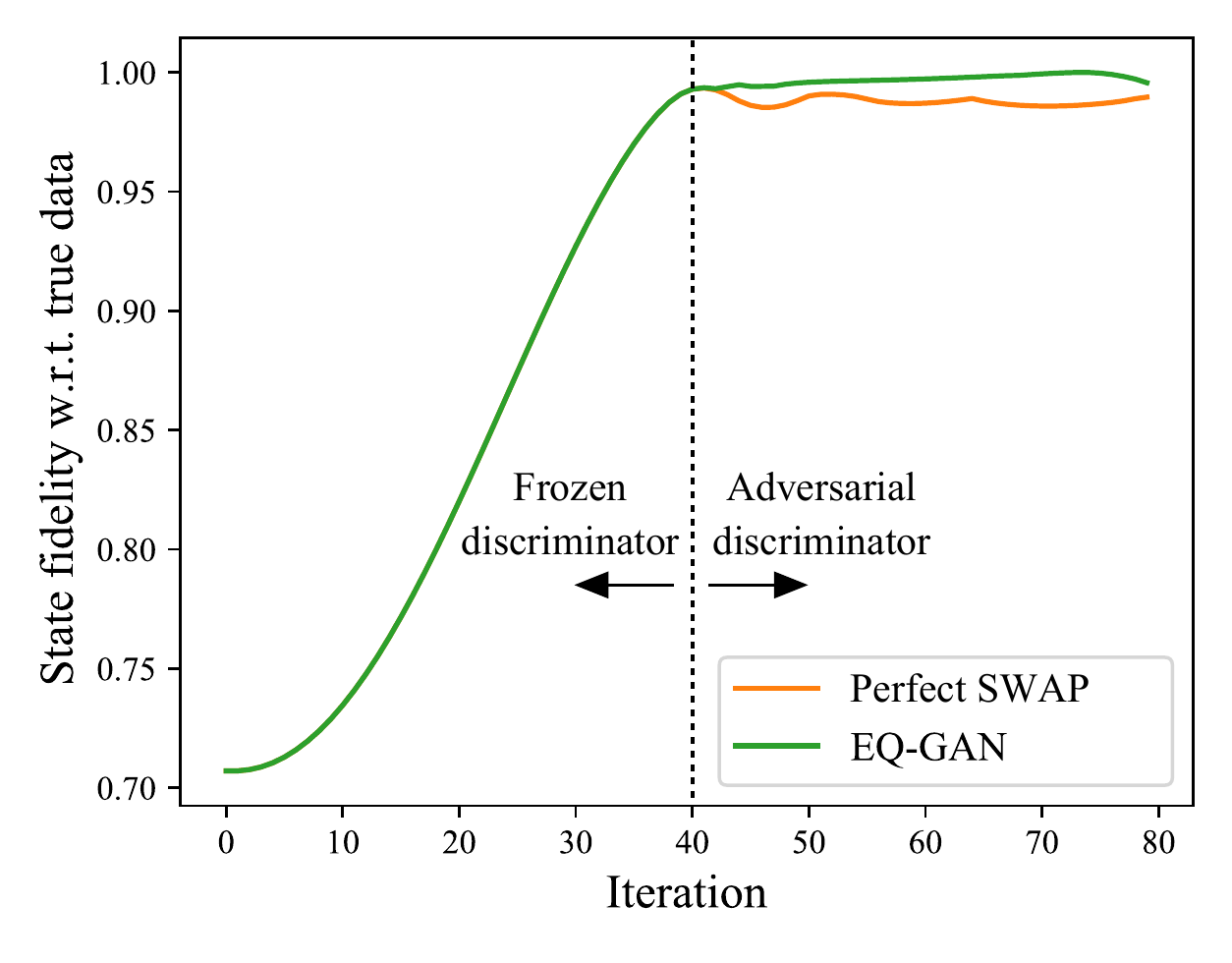}
\caption{Comparison of EQ-GAN and a perfect swap test for a simulated noise model. Normally distributed noise on single-qubit rotations are applied with a systematic bias away from zero, causing the swap test to force convergence to the incorrect state.}
\label{fig:errsim}
\end{center}
\end{figure}

To help ensure stable training of the EQ-GAN architecture, we introduce a training procedure that capitalizes on the fact that the discriminator must converge to a swap test at the optimal Nash equilibrium. Rather than training both the generator and discriminator from the beginning, we pre-train the EQ-GAN in a supervised setting. In the first phase, the discriminator is frozen with the parameters of a perfect swap test, although the unitary $\tilde U(\theta_d^\mathrm{opt})$ may be an imperfect swap test; the generator is trained until the loss converges. In the second phase of training, the discriminator is allowed to vary adversarially against the generator, seeking the parameters $\theta_d^*$. In the context of gate errors, this second phase may yield a unitary closer to a true swap test. The example shown in Fig.~\ref{fig:err} of the main text on a physical quantum devices is replicated in Fig.~\ref{fig:errsim} here, showing the two phases of training and the benefit of an adversarial swap test in the presence of noise.

\section{EQ-GAN Discriminator Architecture}%
\begin{figure}[h!]
\begin{center}
\includegraphics[width=\linewidth]{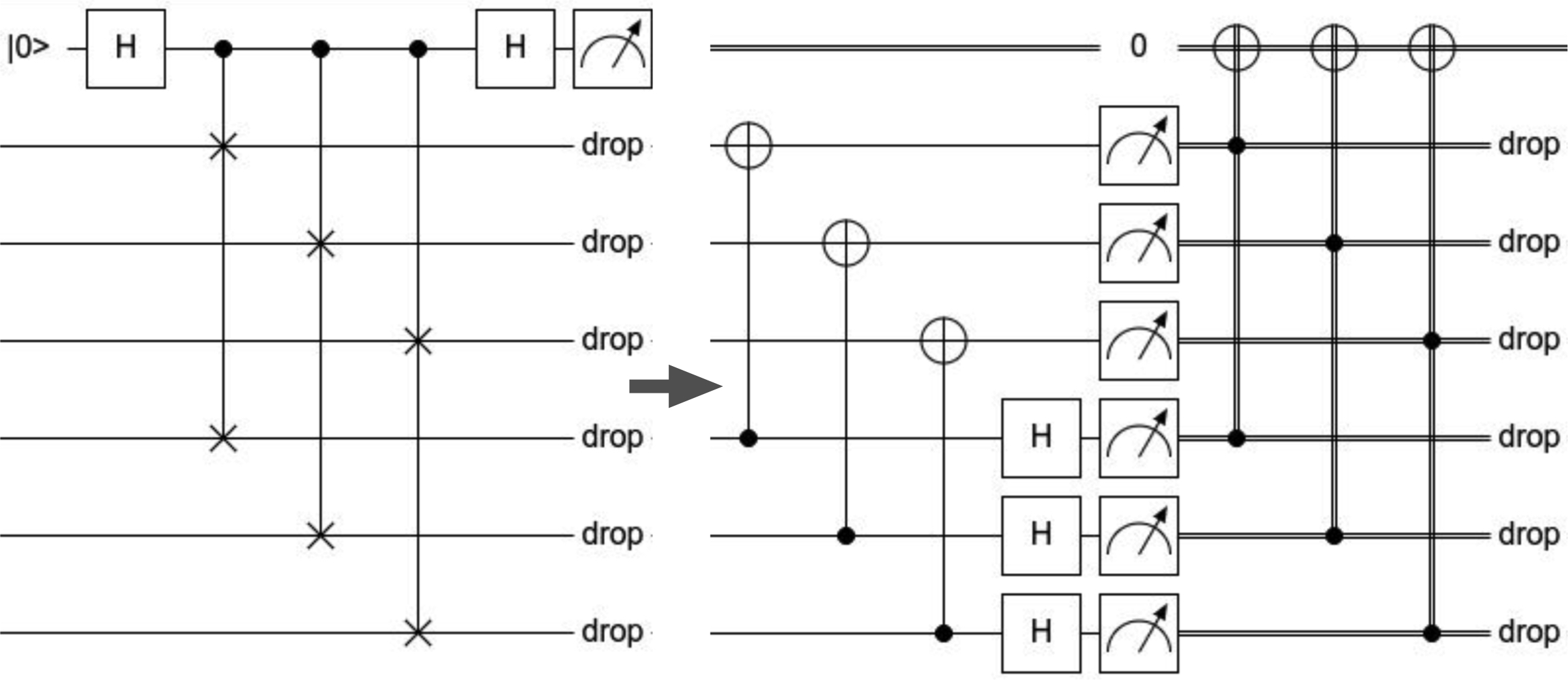}
\caption{Ancilla-free swap test between two 3-qubit states. By rewriting the controlled-swap operation as CNOT and Toffoli gates and replacing computational basis operations with classical post processing, the swap test can be performed with an ancillary classical bit.}
\label{fig:swap}
\end{center}
\end{figure}

We implement an ancilla-free swap test to perform state discrimination (Fig.~\ref{fig:swap}). To evaluate the swap test on a Sycamore quantum device, we decompose each $CNOT$ gate into $(I \otimes H)CZ(I \otimes H)$ operations to use the native $CZ$ gate. As discussed in the main text, the $CZ$ gate has unstable errors that can be effectively modeled with $Z$ rotations by an unknown angle on either qubit. The EQ-GAN formalism can overcome the single-qubit phase error by applying $RZ(\theta)$ gates directly after each $CZ$ operation. During adversarial training, the free angles $\theta$ are optimized with gradient descent to mitigate the two-qubit gate error. Due to the convergence properties provided by the generative adversarial framework, the discriminator provably converges to the best state discriminator possible. This motivates early stopping (as shown in Fig.~\ref{fig:err}) when the discriminator loss indicates that the best state discriminator has been achieved.

\section{QNN Architecture}%
\begin{figure}[h!]
\begin{center}
\includegraphics[width=\linewidth]{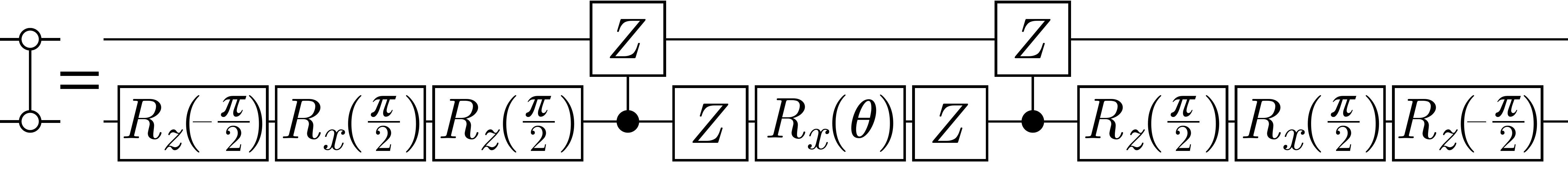}
\caption{Decomposition of the two-qubit entangling gate $G(\theta)$ used in the QNN ansatz (Eq.~\ref{eq:2q}).}
\label{fig:2q}
\end{center}
\end{figure}

We seek to demonstrate an empirical difference in performance between training a quantum neural network (QNN) with individual examples of a classical dataset and with a superposition of data as obtained from a pre-trained EQ-GAN.  To use the native $CZ$ two-qubit gate, we implement a rank-4 entangling gate $G$ given by
\begin{equation}
    \label{eq:2q}
    G(\theta) = \begin{pmatrix}
    1 & 0 & 0 & 0\\
    0 & e^{-i\theta} & 0 & 0\\
    0 & 0 & e^{-i\theta} & 0\\
    0 & 0 & 0 & 1
    \end{pmatrix},
\end{equation}
which is decomposed as shown in Fig.~\ref{fig:2q}.

\begin{figure}[h!]
\begin{center}
\includegraphics[width=\linewidth]{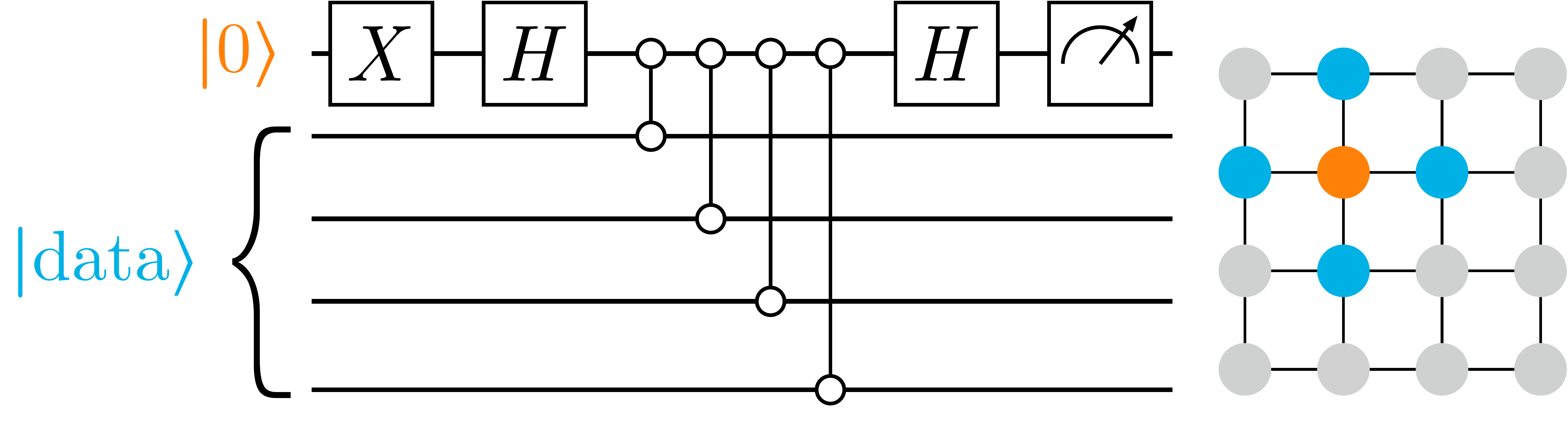}
\caption{Quantum neural network architecture (left) and its corresponding layout on the Sycamore device (right). A four-qubit data state is constructed with the circuits shown in Fig.~\ref{fig:pansatz} and placed in the $\ket{\mathrm{data}}$ state on the blue qubits. A readout qubit (orange) performs parameterized two-qubit interactions shown in Fig.~\ref{fig:2q}.}
\label{fig:qnn}
\end{center}
\end{figure}

Due to the planar connectivity of a Sycamore quantum device, we implement the QNN shown in Fig.~\ref{fig:qnn} with a four-qubit data state. 

The QNN is trained in two ways: it is either trained via \emph{sampling} (shown one training example each iteration, as in Ref.~\cite{farhi2018classification}) or via \emph{superposition} (shown a superposition over an entire class each iteration). As discussed in the main text, the superposition methodology does not use an exact superposition of the training dataset. Instead, it uses a shallow approximation obtained by pre-training an EQ-GAN. We prepare a symmetric concatenation of exponential functions to approximate a peak with minimal circuit depth. In comparison, preparing a Gaussian distribution over $n$ qubits requires $(n-1)$-controlled rotations, which must be decomposed into $2^{n-1}$ $CZ$ gates to use the native gate basis (see Fig. 10 of~\cite{PhysRevA.102.012612}); additional swap operations are required to prepare the state on a planar architecture. Given the empirical dataset, we may also prepare an exact superposition of the data following a state preparation procedure such as that proposed in Ref.~\cite{aaronson2016complexity}. However, this also requires $n$-controlled rotations, leading to an exponential dependence in the number of qubits. All three versions of the QRAM are shown in Fig.~\ref{fig:ansatzes}.

To ensure a fair comparison, we permit an equal number of queries to the quantum device. Consequently, for $N=60$ examples with 30 examples per class, training via sampling is performed for 1 epoch with 60 corresponding to 60 iterations performed on the quantum device. However, training via superposition evaluates the superposition of each class 30 times (since there are two classes), also accessing the quantum device for 60 iterations.

Additionally, Bayesian optimization is used to tune different learning rates for the sampling and superposition methodologies. In simulation, we optimize over Adam learning rates from $10^{-4}$ to $10^{-1}$ with 10 random parameter tries and 40 evaluations of the Gaussian process estimator. For each parameter query, the output of the QNN is averaged over 10 trials to reduce any statistical fluctuations. QNNs using the final learning rates ($10^{-3.93}$ for sampling and $10^{-1.83}$ for superposition) are then evaluated over 50 trials to obtain the final performance reported in Table~\ref{tab:res} with computed standard deviations.

\section{Adversarial Learning Without Errors}%
\begin{figure}[h!]
\centering
\includegraphics[width=0.8\linewidth]{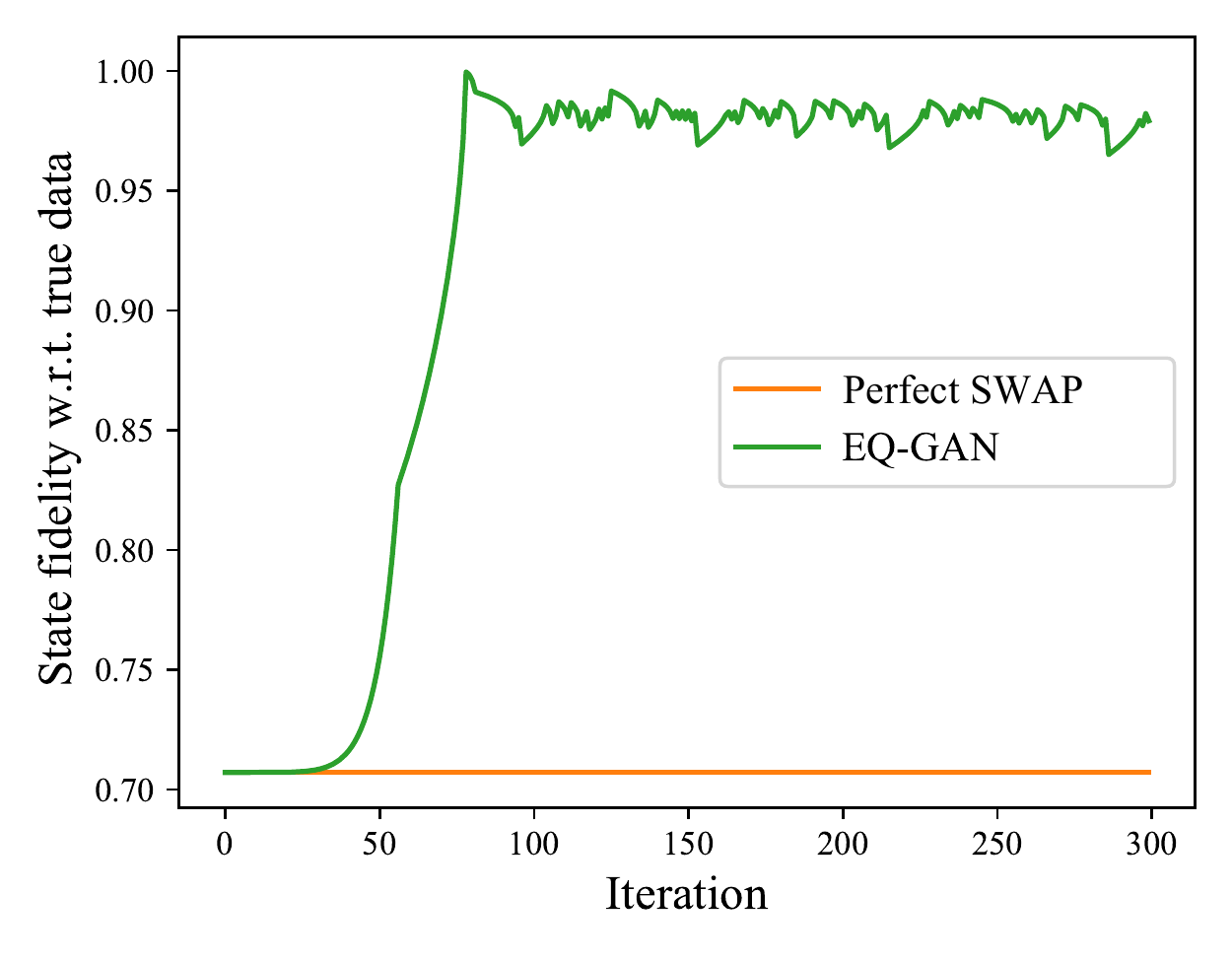}
\caption{Demonstration of a vanishing gradient for a perfect swap test and convergence for the EQ-GAN. While the swap test cannot be trained by gradient descent, the EQ-GAN achieves a state overlap of 0.97.}
\label{fig:vanish}
\end{figure}

While the main text discusses the benefit of adversarial learning in the noisy case, i.e. the automatic suppression of device errors, we provide additional motivation here for using adversarial learning in the \emph{noiseless} case. In particular, we construct an example for which a perfect swap test fails and adversarial learning successfully generates the true data state (Fig.~\ref{fig:vanish}).

\begin{figure}[h!]
\centering
\includegraphics[width=0.8\linewidth]{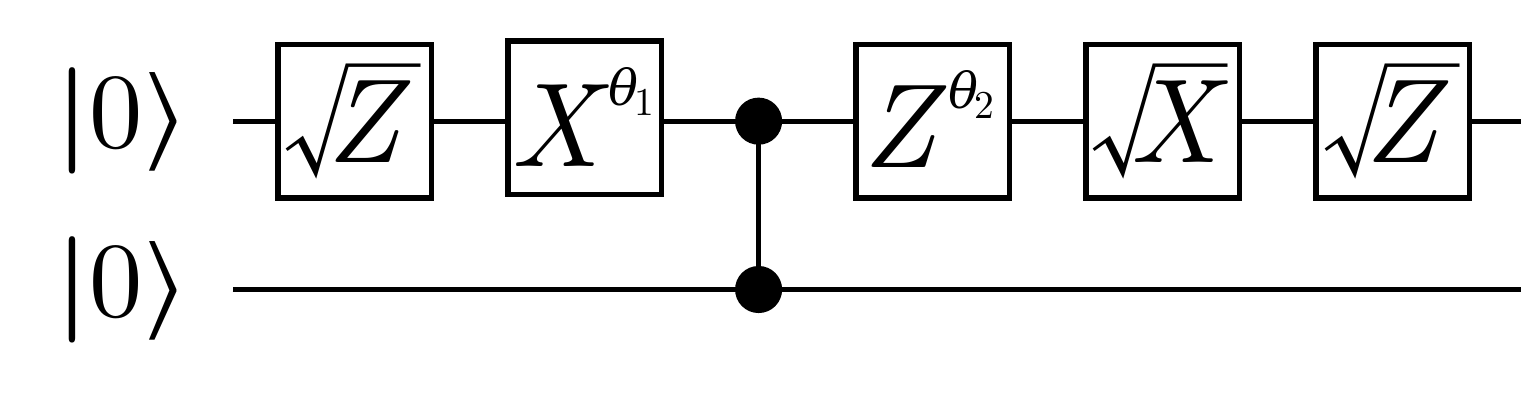}
\caption{Generator and data circuit with a vanishing gradient given data defined by $X$ and $Z$ rotations of $\pi/2$ and a generator initialized with zero angles.}
\label{fig:vanishcircuit}
\end{figure}

\begin{figure*}[t]
\centering
\includegraphics[width=\linewidth]{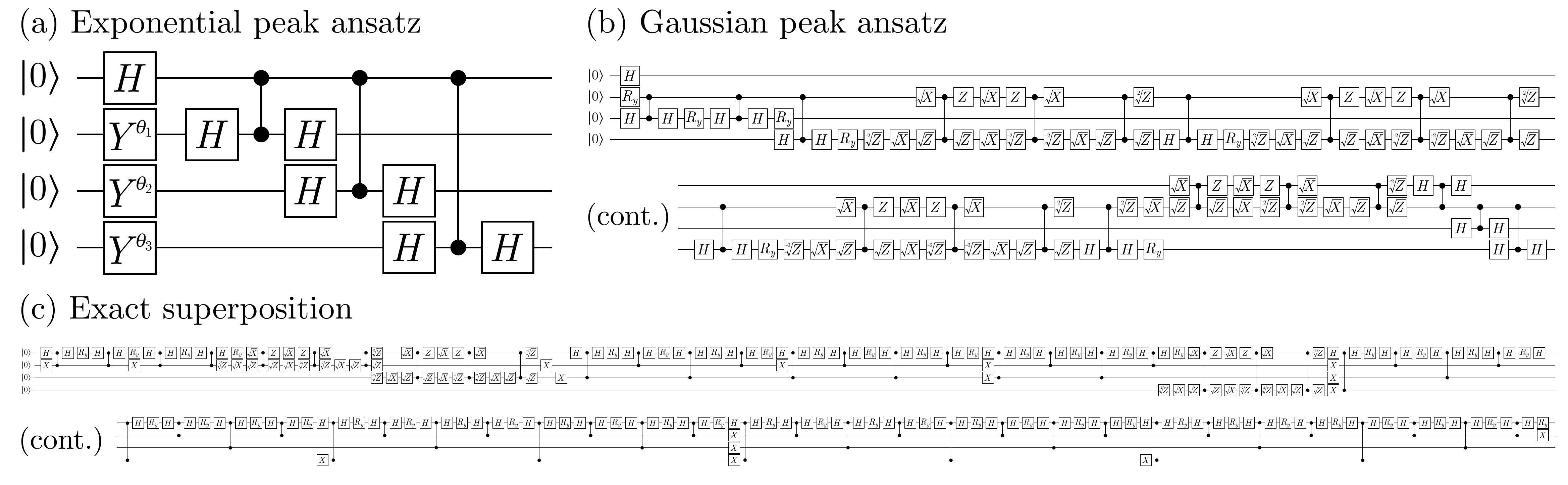}
\caption{QRAM ansatzes for $n=4$ qubits in planar connectivity with (a) exponential peaks (3 two-qubit gates), (b) Gaussian distribution (21 two-qubit gates), and (c) exact superposition (57 two-qubit gates). We adopt ansatz (a) because circuit depth scales polynomially for a QRAM with $n$ qubits, while (b) and (c) scale exponentially with $n$.}
\label{fig:ansatzes}
\end{figure*}

Given the generator ansatz shown in Fig.~\ref{fig:vanishcircuit}, define the data state to have angles $\alpha_0 = \beta_0 = \pi/2$ for corresponding rotations $R_x(\alpha_0), R_z(\beta_0)$. The generator then optimizes angles $\alpha, \beta$ towards achieving full state overlap. In general, the gradient of the state overlap is $\frac{\pi}{4}\sqrt{2 - 2\cos(2 \pi \alpha)\cos(2\pi \beta)}$. By initializing the generator with $\alpha = \beta = 0$, the gradient and all higher derivatives of the overlap vanish. Since a noiseless supervised learning approach with a perfect swap test can only evaluate the gradient of a state overlap measurement, gradient descent will fail to converge to the correct values.

On the other hand, by allowing the discriminator to change, the issue of a vanishing gradient is circumvented and the generator learns the data state (Fig.~\ref{fig:vanish}). For simplicity, we use the same discriminator architecture as that used for suppressing errors. Parameters are optimized with vanilla stochastic gradient descent. The EQ-GAN learning rate schedule is manually tuned, and we verify that no selection of learning rate allows the perfect swap test9 to converge.

\section{Hybrid EQ-GAN for Mixed-State}
\begin{figure}[h!]
\begin{center}
\includegraphics[width=1\linewidth]{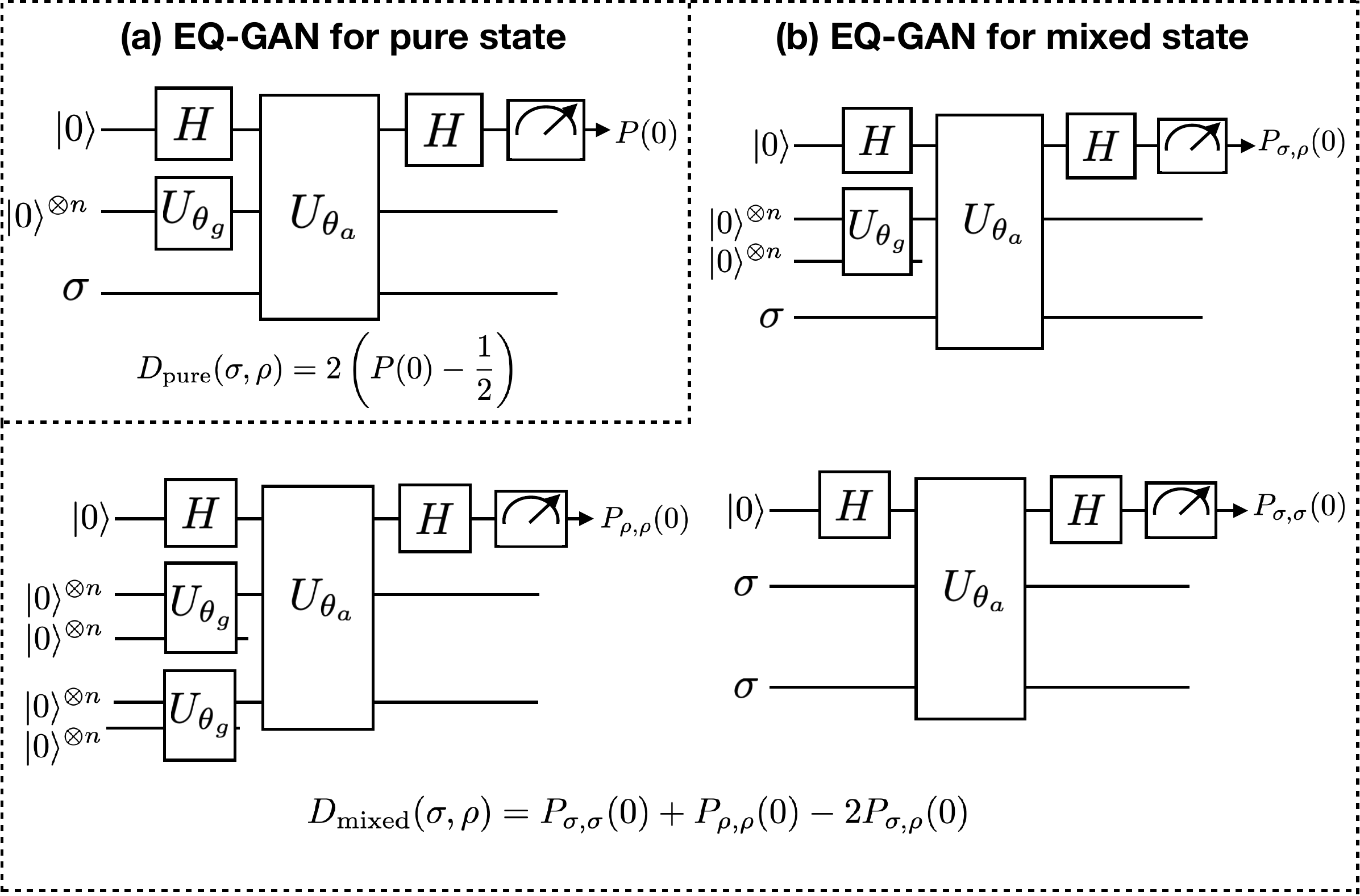}
\caption{Diagram for EQ-GAN architecture based on quantum swap tests. (a) EQ-GAN for learning how to generate pure-state quantum data. (b) EQ-GAN for learning how to generate mixed-state quantum data. The true input data is represented by $\sigma$, and the fake input data $\rho$ is prepared by a unitary circuit $U_{\theta_g}$. The discriminator QNN realizes a unitary transformation represented by $U_{\vec{\theta}_a}$ jointly on the true data, fake data and the ancillary qubit. Measurement on the ancillary qubit is used for the cost function similarly to the EQ-GAN defined in the main text.
\label{QGANsfig}}
\end{center}
\end{figure}

While classical GANs use a random latent vector to generate fake data, the quantum GAN proposed here and in the existing literature does not require any such random input. This comes with a price, especially when our goal is to learn quantum data in a mixed state. As shown in Fig.~\ref{QGANsfig}, a factor of two overhead in the number of qubits are needed for mixed-state learning based on  Choi's theorem.

A closer look at the mathematical nature of a mixed state points us to a more efficient representation through a hybrid classical-quantum network. A mixed state represented in the most generic form $\rho = \sum_{i=1}^{2^n} P_i \ket{\psi_i}\bra{\psi_i}$ is specified by a classical probability distribution $\{P_i\}$ over $2^n$ discrete variables corresponding to the set of quantum states $\{\ket{\psi_i}\}$ that diagonalize the density matrix. Naturally, one can efficiently represent the classical part of this representation, $\{P_i\}$, with a classical neural network, while a quantum circuit prepares the state $\ket{\psi_i}$ given parameter set $\theta_{g_i}$. In this way, we will be able to output a probabilistic mixture of the quantum state by sampling from $\{P_i\}$ and then prepare the associated state.
This obviates the possible double exponential overhead in learning the full quantum channel that transform a fixed initial state to the desired mixed state, as illustrated in Fig.~\ref{QGANsfig}.

\end{document}